\documentclass{iopconfser}

\usepackage{amsmath,graphicx,slashed,booktabs,placeins}
\usepackage[sorting=none]{biblatex}
\usepackage[labelfont=bf]{caption}

\newcommand{\brac}[1]{\left(#1\right)}
\newcommand{\sqbrac}[1]{\left[#1\right]}

\begin{document}

\title{New Isobar Models for $K^+\Lambda$ Electroproduction}

\author{Jovan Alfian Djaja$^{\dagger}$ and Terry Mart\orcid{0000-0003-4628-2245}$^{\ddagger}$}

\affil{Departemen Fisika, FMIPA, Universitas Indonesia, Depok 16424, Indonesia}

\email{$^\dagger$jovan.alfian@ui.ac.id, $^\ddagger$terry.mart@sci.ui.ac.id}

\begin{abstract}
The electroproduction of kaon on proton has been studied using two covariant isobar models. The models incorporate propagators and vertex factors developed in our previous works. In total, the current study includes 26 nucleon resonances and 20 hyperon resonances, with spins up to $13/2$. In the electromagnetic vertices, we consider two alternative electromagnetic form factors alongside the commonly used dipole model. The unknown parameters in the models, such as longitudinal coupling constants and form factor cutoffs, are determined by fitting the calculated observables to nearly 2000 experimental data points. The resulting models demonstrate satisfactory consistency with the available experimental data.
\end{abstract}

\section{Introduction}
The electroproduction of kaon on the proton, $e+p\to e'+K^++\Lambda$, has been investigated for decades \cite{Levy:1973zz,Steininger:1996xw}. More recently, this process has been revisited using a covariant isobar model \cite{Sakinah:2019jwd}, which incorporates nucleon resonances with spins up to $9/2$. In contrast, kaon photoproduction $\gamma +p\to K^++\Lambda$ has been studied more extensively \cite{Thom:1966rm,Adelseck:1985scp,Shinmura:1994pz,Lu:1995bk,Cheoun:1996kn,Janssen:2001wk,Han:1999ck}, owing to its relative simplicity. For example, using an isobar model originally developed for electroproduction, the photoproduction process has been successfully analyzed by including nucleon resonances with spins as high as $13/2$ \cite{Luthfiyah:2021yqe}. However, the electroproduction process offers a unique advantage over photoproduction: the finite momentum of the virtual photon, i.e., $k^2\neq 0$, allows access to the longitudinal properties of hadrons, specifically their electromagnetic form factors. These form factors have been of particular interest for nearly a century, as they offer insights into the internal charge distribution of hadrons and, consequently, their electromagnetic radii. Notably, over the past decade, the so-called "proton radius puzzle" has emerged as one of the most actively debated topics in this context \cite{Ridwan:2023ome,Mart:2013gfa}. Furthermore, the electroproduction process provides access to the generalized Gerasimov$-$Drell$-$Hearn (GDH) sum rule, which connects the nucleon’s ground state to its excited states. As such, it serves as a valuable tool for probing the internal structure of the nucleon \cite{Drechsel:2000ct,Mart:2008zz,Sumowidagdo:1999yp}. 

Although various mechanisms, such as the quark model \cite{Onley,Li:1995si} and partial wave analysis \cite{Mart:2006dk,Mart:2017xtf}, have been developed to explain kaon production, the isobar model \cite{Petrellis:2024ybj, Skoupil:2018vdh,Maxwell:2012zz,Borasoy:2007ku,Wang:2024byt,Mai:2023cbp} offers a particularly simple and efficient approach. Moreover, it can be readily applied in a range of contexts, including nuclear environments and studies of form factors and isospin symmetries.
For a detailed account of recent advances in kaon photo- and electroproduction, readers are encouraged to consult Ref.~\cite{Mart:2021svd}.

In this work, our aim is to extend the isobar model for the photoproduction process that includes nucleon resonances with spins up to $13/2$ to the finite $Q^2$ region, i.e., to electroproduction. In addition, we also investigate the impact of different choices for the electromagnetic form factors of nucleon and hyperon resonances on the model predictions. Specifically, we employ two new parameterizations for these form factors. 
For the kaon resonances $K^*$ and $K_1$, the standard monopole form factor is used. The model is then fitted to a dataset comprising 1953 experimental points to extract unknown parameters, such as electromagnetic and hadronic coupling constants, as well as form factor cutoffs. The experimental data are taken from measurements by the CLAS Collaboration at Jefferson Lab \cite{Ambrozewicz2007,CLAS:2022yzd}. It is also important to note that, although very limited, new electroproduction data have been provided by the Jefferson Lab Hypernuclear Collaboration \cite{JLabHypernuclear:2024mgx}.


\section{Kinematics}
The electroproduction of $K^+\Lambda$ on the proton can be written as an inelastic scattering process
    \begin{equation}
        e(k_i)+p(p)\to e'(k_f)+K^+(q)+\Lambda(p_\Lambda),
    \end{equation}
where the four-momentum of each particle is indicated in the parentheses. This process can also be regarded as a virtual photoproduction process
    \begin{equation}
        \gamma_\text{v}(k)+p(p)\to K^+(q)+\Lambda(p_\Lambda),
    \end{equation}
where $k\equiv k_i-k_f$ is the four-momentum of the virtual photon. The invariant amplitudes obtained from the relevant Feynman diagrams can be decomposed into six gauge- and Lorentz-invariant matrices $M_i$, 
    \begin{equation}
        \mathcal{M}=\bar{u}_\Lambda \sum_{i=1}^6A_i(s,t,u,k^2)M_iu_p,
        \label{eq:A_iM_i}
    \end{equation}
where $\bar{u}_\Lambda\equiv \bar{u}(p_\Lambda)$ and $u_p\equiv u(p)$ are the spinors of the $\Lambda$ and proton, respectively. The Mandelstam variables, $s$, $t$, and $u$ are defined by 
    \begin{equation}
        s=(k+p)^2,\ \ t=(k-q)^2,\ \ u=(k-p_\Lambda)^2
    \end{equation}
while the gauge- and Lorentz- invariant matrices $M_i$ are given by \cite{Luthfiyah:2021yqe,Clymton:2021wof,Mart:2019mtq,Mart:2015jof}
    \begin{align}
	    M_1&=\frac{1}{2}\gamma_5\brac{\slashed{\epsilon}\slashed{k}-\slashed{k}\slashed{\epsilon}},               \\
	    M_2&=\gamma_5\sqbrac{\brac{2q-k}\cdot \epsilon P\cdot k-\brac{2q-k}\cdot k P\cdot \epsilon}, \\
	    M_3&=\gamma_5\sqbrac{\brac{q\cdot k} \slashed{\epsilon}-\brac{q\cdot\epsilon}\slashed{k}},               \\
	    M_4&=i\varepsilon_{\mu\nu\rho\sigma}\gamma^\mu q^\nu\epsilon^\rho k^\sigma,               \\
	    M_5&=\gamma_5\sqbrac{\brac{q\cdot \epsilon}k^2-\brac{q\cdot k}\brac{k\cdot\epsilon}},       \\
	    M_6&=\gamma_5\sqbrac{\brac{k\cdot\epsilon}\slashed{k}-k^2\slashed{\epsilon}}, 
    \end{align}
where $P\equiv (p+p_\Lambda)/2$ and $\varepsilon_{\mu\nu\rho\sigma}$ is the four-dimensional Levi-Civita tensor. Note that in the literature, the definitions of $M_i$'s are not unique. All observables measured in experiments can be calculated analytically using $A_i$ extracted from Eq.~(\ref{eq:A_iM_i}), after accounting for all contributions from the intermediate state of the Born and resonance terms in the amplitude ${\cal M}$. 

\section{Electromagnetic Form Factors}
As a starting point, we use the standard dipole electromagnetic form factor for baryon (i.e., nucleon and hyperon) resonances,  
\begin{equation}
    F_{\rm D}(Q^2)=\left(1+\frac{Q^2}{\Lambda^2}\right)^{-2}.
\end{equation}
Furthermore, we also use two different form factor models given by
\begin{align}
    &F^{B^*}_1(Q^2)=F_{\rm D}(Q^2)\,\exp(-a_1Q^2), \label{Eq.FF1} \\[0.25cm]
    &F^{B^*}_2(Q^2)=F_{\rm D}(Q^2)\,\text{sech}^2(a_2Q^2)\label{Eq.FF2},
\end{align}
where $Q^2\equiv -k^2$ and the free parameters $\Lambda$ and $a_i$ are determined by fitting the model to the available experimental data. 

For kaon resonances $K^*$ and $K_1$, we employ the commonly used monopole form factor
\begin{equation}
    F_{K^*}(Q^2)=\left(1+\frac{Q^2}{\Lambda_{K^*}^2}\right)^{-1} \hspace{0.5cm} \text{and} \hspace{0.5cm} F_{K_1}(Q^2)=\left(1+\frac{Q^2}{\Lambda_{K_1}^2}\right)^{-1},
\end{equation}
where the cutoff masses $\Lambda_i$ are also determined from fitting the observables to the experimental data.

\section{Observables}
The measured differential cross section for kaon electroproduction in an experiment is given by 
\begin{equation}
    \frac{d\sigma}{d\Omega_{e'}dE_{e'}d\Omega_K}= \Gamma\,\frac{d\sigma_v}{d\Omega_K},
\end{equation}
where $\Gamma$ is the flux of virtual photon field~\cite{Knochlein:1995qz}. The virtual differential cross section ${d\sigma_v}/{d\Omega_K}$ can be written in terms of the structure functions $\sigma_i\equiv{d\sigma_i}/{d\Omega_K}$, where $i=$ T, L, U, TT, LT, and LT' refer to transversely unpolarized, longitudinally polarized, unpolarized, transversely polarized, and interference terms between longitudinally and transversely polarized terms, respectively \cite{Sakinah:2019jwd}. Explicitly, it is written as \cite{Mart:2010ch}
\begin{equation}
    \frac{d\sigma_v}{d\Omega_K} = \frac{d\sigma_\mathrm{T}}{d\Omega_K} + \epsilon \frac{d\sigma_\mathrm{L}}{d\Omega_K} + \epsilon\frac{d\sigma_\mathrm{TT}}{d\Omega_K}\cos{2\Phi}+ \sqrt{\epsilon(1+\epsilon)}\frac{d\sigma_\mathrm{LT}}{d\Omega_K}\cos{\Phi}+h\sqrt{\epsilon(1-\epsilon)} \frac{d\sigma_\mathrm{LT'}}{d\Omega_K}\sin{\Phi},
\end{equation}
where $\epsilon$ measures the transverse polarization of the virtual photon and $h$ is the helicity of the electron beam \cite{Knochlein:1995qz,Mart:2010ch}. In this work, only the first three structure functions will be discussed. Additionally, we have experimental data for the averaged beam-recoil transferred polarizations $P'_{x'}$ and $P'_x$ defined as \cite{CLAS:2022yzd}
\begin{eqnarray}
  P'_{x'}(\cos\theta) 
  &=& \frac{\displaystyle 
  \int dW \int dQ^2\, K_{\rm f}
  \sqrt{1-\epsilon^2}\, 
  R_{\rm TT'}^{x'0}(Q^2,W,\cos\theta)}{\displaystyle \int 
  dW\int dQ^2\, K_{\rm f}\left\{ R_{\rm T}^{00}(Q^2,W,\cos\theta)
    +\epsilon R_{\rm L}^{00}(Q^2,W,\cos\theta)
  \right\}}\,,
  \label{eq:RF1}
\end{eqnarray}
and
\begin{eqnarray}
  P'_{x}(\cos\theta) &=& \frac{\displaystyle 
  \int dW\int dQ^2\, K_{\rm f}
  {\textstyle\frac{1}{2}}\sqrt{\epsilon(1-\epsilon)}\, \left(
  R_{\rm LT'}^{x'0}\cos\theta-R_{\rm LT'}^{y'0}+R_{\rm LT'}^{z'0}\sin\theta
  \right)}{\displaystyle \int dW \int dQ^2\, K_{\rm f}\left\{ R_{\rm T}^{00}(Q^2,W,\cos\theta)
    +\epsilon R_{\rm L}^{00}(Q^2,W,\cos\theta)
  \right\}}\,,
  \label{eq:RF2}
\end{eqnarray}
where the response functions $R_i^{\mu\nu}(Q^2,W,\cos\theta)$ and the notation for variables are given, e.g., in Ref.~\cite{Knochlein:1995qz}.

\begin{table}[t]
	\centering
	\caption{Extracted parameters from the photoproduction process used in the present work. The notation of parameters is explained in our previous work \cite{Mart:2019mtq,Mart:2015jof}.}
	\begin{tabular}{c @{\hskip 0.5cm} r @{\hskip 1.5cm} c @{\hskip 0.5cm} r}
		\toprule
		Parameter   &     Value & Parameter & Value \\
		\midrule
		$g_{K\Lambda N}/\sqrt{4\pi}$   & $-3.000$ & $G^T_{K_1}/4\pi$ & $4.511$ \\[0.25cm]
		$g_{K\Sigma N}/\sqrt{4\pi}$ &  $0.900$ & $\Lambda_\text{B}$ (GeV)& $0.700$   \\[0.25cm]
		$G^V_{K^*}/4\pi$ & $0.171$ & 		$\Lambda_\text{R}$ (GeV)& $1.154$ \\[0.25cm]
		$G^T_{K^*}/4\pi$ & $0.182$ & 		$\theta_\text{had}$ (deg)& $114.4$\\[0.25cm]
		$G^V_{K_1}/4\pi$ & $0.605$ & 		$\phi_\text{had}$ (deg)& $0.000$ \\
		\bottomrule
	\end{tabular}
	\label{T.Parameter Background}
\end{table}
\FloatBarrier

\section{Results and Discussion}
In the following discussion, Model 1 and Model 2 refer to the isobar models that employ Eqs.~\eqref{Eq.FF1} and \eqref{Eq.FF2}, respectively, for their electromagnetic form factors. The coupling constants extracted from the fitting process are listed in Tables~\ref{T.Parameter Background} and \ref{tab:parameter}. Table~\ref{T.Parameter Background} shows that the Born coupling constants are constrained to the SU(3) predictions. Additionally, from Table \ref{tab:parameter}  it is evident that Models 1 and 2 yield lower values of $\chi^2/N$ compared to the model employing a standard dipole form factor. This improvement is mathematically expected, as the inclusion of additional parameters $a_i$ in Eqs.~(\ref{Eq.FF1}) and (\ref{Eq.FF2}) enhances the models' flexibility in reproducing the experimental data.
Physically, this suggests that the $Q^2$ dependence of the observables decreases more rapidly than that captured by a dipole form factor, implying that the recent CLAS data suggest a more compact charge distribution for the baryon resonances.

\begin{table}[hbt!]
	\centering
	\caption{The extracted longitudinal coupling constants of nucleon resonances. The parameters were obtained by fitting the models to 1953 experimental data points. The number of parameters and $\chi^2/N$ of each model are shown below.}
	\begin{tabular}{l @{\hskip 0.75cm} r @{\hskip 0.75cm} r @{\hskip 0.75cm} r}
		\toprule
		Parameter &  Dipole & Model 1 & Model 2\\
		\midrule
		$G_{N\brac{1440}}$  & $-4.92840$ & $-4.67770$ & $-4.61450$  \\[0.1cm]
		$G_{N\brac{1535}}$  & $-0.11859$ & $-0.11481$ & $-0.09409$  \\[0.1cm]
		$G_{N\brac{1650}}$  & $-0.00418$ & $-0.00630$ & $0.63124$   \\[0.1cm]
		$G_{N\brac{1710}}$  & $0.47025$  & $0.33743$  & $1.46520$   \\[0.1cm]
		$G_{N\brac{1880}}$  & $0.00087$  & $0.00026$  & $0.81793$   \\[0.1cm]
		$G_{N\brac{1895}}$  & $-0.00268$ & $-0.00426$ & $-0.00220$  \\[0.1cm]
		$G_{N\brac{2100}}$  & $0.05802$  & $0.09904$  & $0.11529$   \\[0.1cm]
		$G_{N\brac{2300}}$  & $6.70740$  & $6.77580$  & $6.41590$   \\
		\midrule
		$\chi^2$        & $4552$ & $4222$ & $4234$ \\[0.1cm]
		$N_\text{par}$  & $117$  & $240$  & $240$ \\[0.1cm]
		$N_\text{data}$ & $1953$ & $1953$ & $1953$ \\[0.1cm]
		$\chi^2/N$      & $2.33$ & $2.16$ & $2.17$ \\
		\bottomrule
	\end{tabular}
	\label{tab:parameter}
\end{table}

\FloatBarrier

The results of the present work are compared with experimental data taken from the CLAS Collaboration \cite{Ambrozewicz2007,CLAS:2022yzd} in Figs~\ref{fig:sigma_u}-\ref{carman}. For completeness, we also display the {\small KAON-MAID} prediction \cite{kaonmaid} as an additional point of comparison.

\subsection{Differential Cross Section}

\begin{figure}
    \centering
    \includegraphics[scale=1.05]{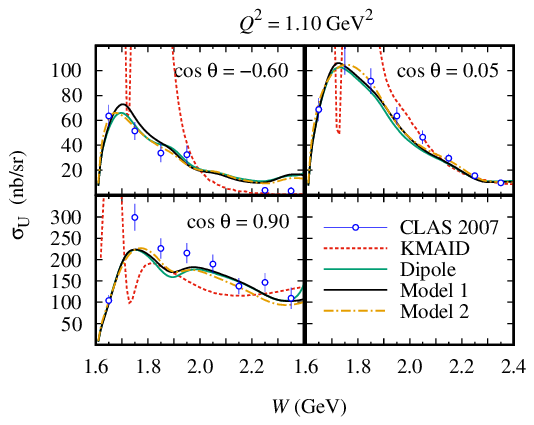}
    \caption{Structure functions $\sigma_{\rm U}\equiv {d\sigma_\mathrm{T}}/{d\Omega_K} + \epsilon {d\sigma_\mathrm{L}}/{d\Omega_K}$ calculated using different models as a function of c.m. energy $W$. Experimental data are taken from Ref.~\cite{Ambrozewicz2007}.}
    \label{fig:sigma_u}
\end{figure}

The unpolarized structure function $\sigma_\mathrm{U} = \sigma_\mathrm{T}+\epsilon\sigma_\mathrm{L}$ as a function of the total c.m. energy $W$ is shown in Fig.~\ref{fig:sigma_u}. In this case, the three models agree well with the experimental data, except for the forward angles. For the forward angle $\cos{\theta}=0.90$, the models slightly underpredict the structure function around $W\approx 1.75\ \mathrm{GeV}$. We also observe that all models reproduce the energy dependence of the unpolarized structure function equally well. There is a clear difference between the present result and the prediction of {\small KAON-MAID}. In the present work, all models exhibit a significant contribution of $P_{11}(1710)$ near $W=1.7$ GeV. This is in contrast to the prediction of {\small KAON-MAID}, which shows very strong contributions of the nucleon resonances $S_{11}(1650)$ and $D_{13}(1895)$.

\begin{figure}
    \centering
    \includegraphics[scale=1.05]{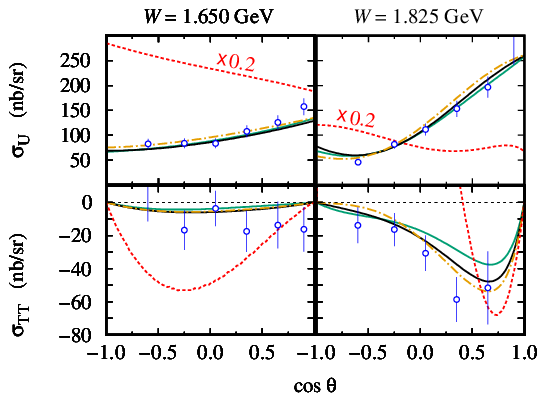}
    \caption{Angular distributions of the structure functions $\sigma_{\rm U}\equiv {d\sigma_\mathrm{T}}/{d\Omega_K} + \epsilon {d\sigma_\mathrm{L}}/{d\Omega_K}$ and $\sigma_\mathrm{TT}\equiv d\sigma_{\mathrm{TT}}/d\Omega_K$ predicted by different models in the present work compared with the prediction of {\small KAON-MAID} and experimental data from the CLAS Collaboration \cite{Ambrozewicz2007} for two different values of the total c.m. energy $W$ at $Q^2=0.65\ \text{GeV}^2$. Notation of the curves is as in Fig.~\ref{fig:sigma_u}. For comparison purposes, the {\small KAON-MAID} predictions in the upper panels have been reduced by a factor of $0.2$.}
    \label{fig:sigma_u_and_t}
\end{figure}

\begin{figure}
    \centering
    \includegraphics[scale=1.05]{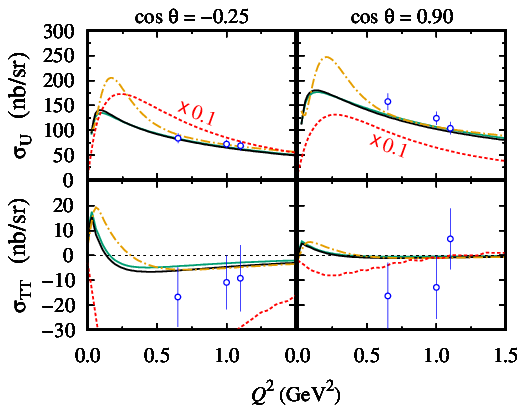}
    \caption{$Q^2$ distributions of the structure functions $\sigma_{\rm U}\equiv {d\sigma_\mathrm{T}}/{d\Omega_K} + \epsilon {d\sigma_\mathrm{L}}/{d\Omega_K}$ and $\sigma_\mathrm{TT}\equiv d\sigma_{\mathrm{TT}}/d\Omega_K$ predicted by different models in the present work compared with the prediction of {\small KAON-MAID} and experimental data from the CLAS Collaboration \cite{Ambrozewicz2007} for two different values of kaon scattering angle $\cos{\theta}$ at $W=1.65\ \text{GeV}$. Notation of the curves is as in Fig.~\ref{fig:sigma_u}. Note that the {\small KAON-MAID} predictions 
    shown in the upper panels have been rescaled by a factor of $0.1$ to fit in the panels.}
    \label{fig:sigma_u_and_tt_Q2}
\end{figure}

We also present the angular distribution of the structure functions $\sigma_\mathrm{U}$ and $\sigma_\mathrm{TT}$ in Figs.~\ref{fig:sigma_u_and_t} and \ref{fig:sigma_u_and_tt_Q2}. As shown in Fig.~\ref{fig:sigma_u_and_t}, all models reproduce the forward peaking of $\sigma_\mathrm{U}$ equally well. However, Models 1 and 2 reproduce the forward peaking of $\sigma_\mathrm{TT}$ better than the Dipole model. This forward peaking characteristic signifies the dominance of $t$-channel exchanges, i.e., kaon resonances exchanges, at forward angles. An interesting difference can be seen in the structure function $\sigma_\mathrm{U}$ predicted by {\small KAON-MAID} compared to that obtained in the present work. While {\small KAON-MAID} predicts a structure function that peaks at backward angles, our results exhibit a forward-peaking behavior. Notably, the present calculations are in better agreement with the experimental data, as clearly demonstrated in the upper panels of Fig.~\ref{fig:sigma_u_and_t}.

The structure functions $\sigma_\mathrm{U}$ and $\sigma_\mathrm{TT}$ as functions of the virtual photon four-momentum transfer squared, $Q^2$, are shown in Fig.~\ref{fig:sigma_u_and_tt_Q2}. All models show good agreement with the experimental data within the reported uncertainties. Notably, Model 2 displays a pronounced peak near the photoproduction point ($Q^2 = 0$) in both $\sigma_\mathrm{U}$ and $\sigma_\mathrm{TT}$, in contrast to the other models. The $Q^2$ dependence of these structure functions reflects a balance between the intrinsic dynamics of the isobar model and the suppression effects introduced by the electromagnetic form factors \cite{JLabHypernuclear:2024mgx}. As demonstrated in a previous study by the CLAS Collaboration \cite{JLabHypernuclear:2024mgx}, this peak structure is primarily attributed to the longitudinal couplings of the nucleon resonances rather than the electromagnetic form factors. Figures~12 and ~13 of Ref.~\cite{JLabHypernuclear:2024mgx} show that the longitudinal term $\epsilon\sigma_\mathrm{L}$ is clearly the dominant term compared to the transverse term $\sigma_\mathrm{T}$ in the formation of the peak structure of $\sigma_\mathrm{U}$. 

\FloatBarrier

\subsection{Double Polarization Observables}

\begin{figure}[hbt]
    \centering
    \includegraphics[scale=1.25]{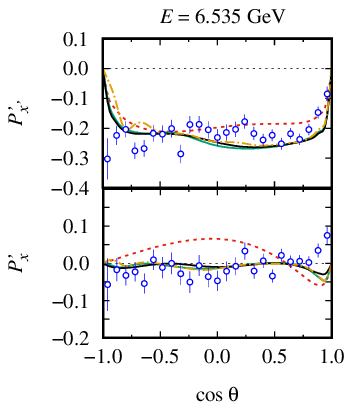}
    \caption{Angular distributions of the beam-recoil transferred polarizations $P'_{x'}$ and $P'_x$ predicted by different models in the present work compared with experimental data from the CLAS Collaboration \cite{CLAS:2022yzd} and the prediction of {\small KAON-MAID} 
    for a beam energy of $E=6.535\ \text{GeV}$. Notation of the curves is as in Fig.~\ref{fig:sigma_u}.}
    \label{carman}
\end{figure}

In addition to the structure functions data, we have also analyzed the beam-recoil transferred polarization $P'_{x'}$ and $P'_x$ measurements from the CLAS Collaboration \cite{CLAS:2022yzd}. In Fig.~\ref{carman}, we display the comparison between our models and the experimental data as a function of $\cos \theta$. We observe that all models can successfully reproduce the overall behavior of $P'_{x'}$ and $P'_x$, which remain approximately constant at $-0.2$ and zero, respectively, across the full range of kaon scattering angles, implying a weak dependence of the observables on the kaon scattering angle. However, none of the models accurately describe $P'_{x'}$ at backward angles, where they tend to rise toward zero instead of maintaining the expected value of $P'_{x'} \approx -0.2$.

The relatively constant behavior of the double polarization observables $P'_{x'}$ and $P'_x$ over the full range of kaon scattering angles is also evident in a previous experiment by the CLAS Collaboration conducted at a lower beam energy \cite{CLAS:2009sbn}. Figures 12 and 13 of Ref.~\cite{CLAS:2009sbn} show that the values of $P'_{x'}$ and $P'_x$ follow a similar pattern to those observed in the present work. Specifically, $P'_{x'}$ remains nearly constant at a negative value before rising toward zero at forward angles, a behavior dictated by angular momentum conservation, which similarly constrains the observable at backward angles. Meanwhile, the values of $P'_{x'}$ remain close to zero, indicating that these observables exhibit only a weak dependence on both the beam energy and the kaon scattering angle.

Finally, it is worth noting that none of the previous analyses, e.g., Ref.~\cite{Mai:2023cbp}, have included the beam-recoil transferred polarization data in their fits. This omission is primarily due to the computational challenges involved, since evaluating these observables requires a double integration of the response functions $R_i^{\mu\nu}(Q^2,W,\cos\theta)$ over $W$ and $Q^2$, as defined in Eqs.~(\ref{eq:RF1}) and (\ref{eq:RF2}). Clearly, this procedure demands substantial computational resources, particularly when performed repeatedly within a data fitting routine.

\FloatBarrier

\section{Conclusions}
We have analyzed recent experimental data for kaon electroproduction $e+p\to e'+K^++\Lambda$ by means of a covariant isobar model. For this purpose, we use new electromagnetic form factors for the nucleon and hyperon resonances in the production amplitude, in addition to the conventional dipole form. In the case of structure functions, the proposed models exhibit good agreement with the experimental data. Furthermore, the angular distribution of the beam-recoil transferred polarizations $P'_{x'}$ and $P'_x$ is well reproduced, in contrast to previous analyses.

\section{Acknowledgements}
This work has been supported by the PUTI Q1 Research Grant from the University of Indonesia (UI) under contract No. NKB 442/UN2.RST/HKP.05.00/2024.

\newpage

\printbibliography

\end{document}